\documentclass[11pt,twoside]{article}

\usepackage{asp2014}

\resetcounters

\aspvoltitle{ASTRONOMICAL DATA ANALYSIS SOFTWARE AND SYSTEMS XXVI}
\aspvolume{in press}
\aspcpryear{ }
\aspvolauthor{ }

\begin{document}

\title{The design strategy of scientific data quality control software for Euclid mission.}


\author{Massimo~Brescia$^1$, Stefano~Cavuoti$^1$, Terje Fredvik$^2$,
  Stein Vidar Hagfors Haugan$^2$, Ghassem Gozaliasl$^{3,6}$, Charles
  Kirkpatrick$^3$, Hannu Kurki-Suonio$^3$, Giuseppe Longo$^4$, Kari
  Nilsson$^{5}$, Martin Wiesmann$^2$
\affil{$^1$INAF OACN - Astronomical Observatory of Capodimonte, Naples, Italy; \email{brescia@na.astro.it}}
\affil{$^2$Institute of Theoretical Astrophysics, University of Oslo, Norway}
\affil{$^3$ Dept. of Physics, University of Helsinki, Finland}
\affil{$^4$ Dept. of Physics University of Naples Federico II, Naples, Italy}
\affil{$^5$ Finnish Centre for Astronomy with ESO (FINCA), University of Turku, Finland}
\affil{$^6$ Helsinki Institute of Physics, University of Helsinki, Finland}
}



\paperauthor{Massimo~Brescia}	{brescia@oacn.inaf.it}{0000-0001-9506-5680}{INAF - Istituto Nazionale di Astrofisica}{OACN - Astronomical Observatory of Capodimonte}{Napoli}{Napoli}{80131}{Italy}
\paperauthor{Stefano~Cavuoti}	{stefano.cavuoti@gmail.com}{0000-0002-3787-4196}{INAF - Istituto Nazionale di Astrofisica}{OACN - Astronomical Observatory of Capodimonte}{Napoli}{Napoli}{80131}{Italy}
\paperauthor{Terje~Fredvik}{terje.fredvik@astro.uio.no}{}{University of Oslo}{Institute of Theoretical Astrophysics}{Oslo}{Oslo}{0316}{Norway}
\paperauthor{Stein~Vidar~Hagfors~Haugan}{}{s.v.h.haugan@astro.uio.no}{University of Oslo}{Institute of Theoretical Astrophysics}{Oslo}{Oslo}{0316}{Norway}
\paperauthor{Martin~Wiesmann}{martin.wiesmann@astro.uio.no}{}{University of Oslo}{Institute of Theoretical Astrophysics}{Oslo}{Oslo}{0316}{Norway}
\paperauthor{Ghassem~Gozaliasl}{ghassem.gozaliasl@helsinki.fi}{}{University of Helsinki}{Dept. of Physics}{Helsinki}{Helsinki}{00560}{Finland}
\paperauthor{Charles~Kirkpatrick}{charles.kirkpatrick@helsinki.fi}{}{University of Helsinki}{Dept. of Physics}{Helsinki}{Helsinki}{00560}{Finland}
\paperauthor{Hannu Kurki-Suonio}{hannu.kurki-suonio@helsinki.fi}{}{University of Helsinki}{Dept. of Physics}{Helsinki}{Helsinki}{00560}{Finland}
\paperauthor{Giuseppe~Longo}{longo@na.infn.it}{University of Naples}{}{Dept. of Physics}{Naples}{Naples}{80126}{Italy}
\paperauthor{Kari~Nilsson}{kari.nilsson@utu.fi}{}{University of Turku}{FINCA - Finnish Centre for Astronomy with ESO}{Turku}{Turku}{21500}{Finland}

\begin{abstract}
The most valuable asset of a space mission like Euclid are the data.
Due to their huge volume, the automatic quality control becomes a
crucial aspect over the entire lifetime of the experiment. Here we
focus on the design strategy for the Science Ground Segment (SGS) Data
Quality Common Tools (DQCT), which has the main role to provide
software solutions to gather, evaluate, and record quality information
about the raw and derived data products from a primarily scientific
perspective. The stakeholders for this system include Consortium
scientists, users of the science data, and the ground segment data
management system itself.  The SGS DQCT will provide a quantitative
basis for evaluating the application of reduction and calibration
reference data (flat-fields, linearity correction, reference catalogs,
etc.), as well as diagnostic tools for quality parameters, flags,
trend analysis diagrams and any other metadata parameter produced by
the pipeline, collected in incremental quality reports specific to
each data level and stored on the Euclid Archive during pipeline
processing.  In a large programme like Euclid, it is prohibitively
expensive to process large amount of data at the pixel level just for
the purpose of quality evaluation. Thus, all measures of quality at
the pixel level are implemented in the individual pipeline stages, and
passed along as metadata in the production. In this sense most of the
tasks related to science data quality are delegated to the pipeline
stages, even though the responsibility for science data quality is
managed at a higher level.  The DQCT subsystem of the SGS is currently
under development, but its path to full realization will likely be
different than that of other subsystems. Primarily because, due to a
high level of parallelism and to the wide pipeline processing
redundancy, for instance the mechanism of double Science Data Center
for each processing function, the data quality tools have not only to
be widely spread over all pipeline segments and data levels, but also
to minimize the occurrences of potential diversity of solutions
implemented for similar functions, ensuring the maximum of coherency
and standardization for quality evaluation and reporting in the SGS.
\end{abstract}

\section{Introduction}

The Euclid Science Ground Segment (SGS, \citealt{laureijs2014}) has
the main role to provide the all the resources required to analyze the
Euclid Data and to derive science data products.  The Euclid SGS is in
charge of the production of the science ready calibrated images and
source catalogues, and all relevant quality control and meta-data
required for the scientific exploitation of the Euclid mission. This,
of course, includes the data from the two instrument channels of
Euclid (VIS and NISP), as well as all complementary external data from
other surveys.

From the data flow point of view, the SGS data processing is composed by four levels:
\begin{itemize}
\item Level 1: Unpacked and checked telemetry and housekeeping data;
\item Level 2: Calibrated and intermediate data produced during the calibrations;
\item Level 3: Final catalogues and pre-defined science data products (3D galaxy power spectra, dark matter power spectra, tomography, high order statistics, mass reconstruction map, photometrin and spectroscopic redshift, etc.) but does not include data analysis beyond the production of catalogues and basic 2-point statistics or cosmological interpretation of data (joint analyses of data, dark energy studies, cosmological parameters, growth and growth rate of structures, galaxy biasing, test gravity, neutrino mass constraints, galaxy clustering, etc...).
\item External Data: Euclid compliant quality-controlled data from existing surveys used for calibrations, photometric redshift, and simulation validations during all the operational phases.
\end{itemize}

The most valuable asset of the mission are the data, and due to their huge  volume, quality control becomes a crucial aspect of all above items, not  only for scientific data produced by the pipeline, but also for telemetry, diagnostics/monitoring/control, and calibration information. Data Quality (DQ) refers to the state of completeness, validity, consistency, timeliness and accuracy that makes data appropriate for a specific use. This means that all kinds of products  (images, tables, text, etc.) have to be quality controlled by checking the right syntax of metadata and columns, detecting missing or out-of-bounds values and detecting any other inconsistencies. DQ common tools are foreseen to be available for all pipeline modules at all processing levels. They should be as general as possible, in order to be shared between different pipeline modules, thus ensuring a full homogeneity.


The DQCT is a system devoted to provide common solutions for the quality controls to be integrated within all the project pipelines. DQCT deals with information concerning quality flags, error estimates, statistics like mean, standard deviation, RMS, S/N, visual/graphical inspection products, such as thumbnails, trend analysis diagrams, histograms, scatter plots, etc., as well as any other metadata parameter produced by the pipeline, collected in quality reports specific to each data level and stored on Euclid archive during pipeline processing.

All the quality tools are based on specifications expected from the scientific organization units, and they would fulfill all common needs at all pipeline flow stages, according to standards defined among all the science teams.

DQCT APIs will also provide tools to collect quality reports at each transition between two data levels, taking also traceability of the previous quality checks along the pipeline into account.

The DQCT team is composed by different Institutes from Finland, Italy and Norway, under the Italian responsibility. During the prototyping and development phases, the produced packages will be tested and validated in collaboration with the Finnish Science Data Center.
The DQCT will be developed by adopting the common development directives defined by System Team in the areas architecture \& design, software development rules and processes, Euclid Archive System, Monitoring \& Control.

\section{Design and Features}

DQCT will provide a quantitative basis for evaluating the application of calibration reference data (flat-fields,
linearity correction, reference catalogs, focal plane illumination model, photometric scale and zero-point, etc.).

DQCT will provide diagnostic tools that, among other things, will facilitate diagnosing problems with  scientific performance and delivered image quality. Analysis of pipeline stage quality problems with DQCT will inform decisions taken by down-stream pipelines and processes of whether to abort or otherwise alter their processing.

The DQCT subsystem is currently under development, but its path to full realization will likely be different in flavor than that of other subsystems. First because, by reflecting historical patterns of prior challenging missions, there is a great deal to learn about Euclid, the best strategies for data processing, new science demands and opportunities that will undoubtedly emerge during operations. Second, because, due to the wide pipeline processing redundancy (the mechanism of double Science Data Center for each processing function) and to a high level of parallelism, the data quality tools has not only to be widely spread over all pipeline segments and data levels, but also to minimize the occurrences of potential diversity of solutions implemented for same functions, ensuring the maximum of coherency and standardization for quality evaluation and reporting in the SGS.

There are many aspects of science data quality that are common among astronomical imaging surveys. Most of these tests can be automated, although in practice most prior surveys depended upon human visual inspection to confirm the computed quality metrics.

For instance common elements of Science Data Quality Assessment for Surveys are: Image artifact flagging: Static bad pixels, cosmic rays, saturation, satellite trails, electronic cross-talk; Image background: Ghost images, scattered light, detector health, moonlight, fringing, sky glow; Delivered image quality: Size of the PSF, PSF shape (e.g., ellipticity), variation of the PSF with position in the focal plane; Astrometric fidelity and stability; Photometric fidelity and stability: Uniformity of photometric depth; dispersion about expected stellar locus in color magnitude diagrams and color-color plots.


During the instrument commissioning, followed by full-up science operations, the DQCTs will contribute to calibrate and assess both stability and repeatability of the science data within a variety of operating conditions in particular for DQCT functions there will be the possibility to correctly set thresholds for quality parameters and flags.

The DQCT system is based on three main tasks: \textit{i} the automated computation and flagging  of off-nominal conditions and quality attributes, by supporting the final assessment of science data quality; \textit{ii} long-term trend analyses, enabling the optimization of data calibration, pipeline processing functions and investigation of quality anomalies; \textit{iii} to avoid redundancy and inhomogeneous calculation of the same quantities.

The Euclid pipelines are required to report problems occurring during processing.
At the end of SGS data flow the result of DQCTs consist of a quality report including all statistics, trend analysis plots, flags and parameters incrementally evaluated from progenitors to final science data.

Furthermore DQCT is responsible to produce all the data that has to be visualized in a tool that runs as part of the Euclid Archive System.  This tool aims to provide a user friendly method of quality information inspections for products and their progenitors. The tool is supposed to be a tailored version of QualityWise \citep{McFarland2013}.

\section{Conclusions}

Over time, the DQCT team will provide tools for advanced search, access, and analysis of archived data products.

For each pipeline of the Euclid SGS DQCT team is supposed to design a specific data model dedicated to list and describe all the parameters and quality flags required to perform the quality controls as well as DQ functions and analyses, particularly suitable to monitor the status of observations.
The final role of the DQCT team will be to provide an incremental quality report tools for advanced search, access, and analysis of archived data products.



\acknowledgements MB and SC acknowledge financial contribution from the agreement ASI/INAF I/023/12/1.
CK and GG were supported by the Academy of Finland grant 257989. KN was supported by Academy of Finland
Grant 295114.

\bibliography{P2-13}  

\end{document}